\documentclass[12pt,preprint]{aastex}
\usepackage{amsmath}
\usepackage{graphicx}
\newcommand{\rin}{R_{\rm in}}

\newcommand{\msun}{\rm M_{\sun}}
\newcommand{\rchi}{\chi^{2}_\nu}

\newcommand{\fsc}{f_{\rm SC}}
\newcommand{\nh}{N_{\rm H}}

\newcommand{\kpc}{\rm kpc}
\newcommand{\kev}{\rm keV}
\newcommand{\keV}{\rm keV}

\newcommand{\cm}{\rm cm}
\newcommand{\km}{\rm km}
\newcommand{\nin}{n_{\rm in}}
\newcommand{\nout}{n_{\rm out}}

\shorttitle{A Simple Comptonization Model}
\shortauthors{Steiner et al.}

\begin{document}

\title{A Simple Comptonization Model}
\author{James F. Steiner, Ramesh Narayan, Jeffrey E. McClintock}
\affil{Harvard-Smithsonian Center for Astrophysics, Cambridge MA 02138}
\email{jsteiner@cfa.harvard.edu}
\and
\author{Ken Ebisawa}
\affil{Institute of Space and Astronautical Science/JAXA, 3-1-1
Yoshinodai, Sagamihara, Kanagawa 229-8510, Japan}

\begin{abstract}

We present an empirical model of Comptonization for fitting the
spectra of X-ray binaries.  This model, {\sc simpl}, has been
developed as a package implemented in XSPEC.  With only two free
parameters, {\sc simpl} is competitive as the simplest empirical model of
Compton scattering.  Unlike other empirical models, such as the
standard power-law model, {\sc simpl} incorporates the basic physics
of Compton scattering of soft photons by energetic coronal electrons.
Using a simulated spectrum, we demonstrate that {\sc simpl} closely
matches the behavior of physical Comptonization models which consider
the effects of optical depth, coronal electron temperature, and
geometry.  We present fits to {\it RXTE} spectra of the black-hole
transient H1743--322 and a {\it BeppoSAX} spectrum of LMC X--3 using
both {\sc simpl} and the standard power-law model.  A comparison of
the results shows that {\sc simpl} gives equally good fits and a
comparable spectral index, while eliminating the troublesome
divergence of the standard power-law model at low energies.
Importantly, {\sc simpl} is completely flexible and can be used self-consistently with
any seed spectrum of photons. 
We show that {\sc simpl} -- unlike the standard power law -- teamed up
with {\sc diskbb} (the standard model of disk accretion) gives results for the
inner-disk radius that are unaffected by strong Comptonization, a result
of great importance for the determination of black hole spin via the
continuum-fitting method.

\end{abstract}


\section{Introduction}

Spectra of X-ray binaries typically consist of a soft (often blackbody
or bremsstrahlung) component and a higher-energy tail component of
emission, which we refer to generically as a ``power law'' throughout
this work.  The origin of the power-law component in both neutron-star
and black-hole systems is widely attributed to Compton up-scattering
of soft photons by coronal electrons \citep[][hereafter
RM06]{White_1995,RR_JEM_review_2006}.  This component is present in
the spectra of essentially all X-ray binaries, and it occurs for a
wide range of physical conditions.

The tail emission is generally modeled by adding a simple power-law
component to the spectrum, e.g., via the model {\sc powerlaw} in the
widely used fitting package XSPEC \citep{XSPEC}.  A few of the many
applications where   power-law models are employed include: modeling
the thermal continuum \citep{Shafee_spin} or the
relativistically-broadened Fe K line \citep{Miller_GX339} in order to
obtain estimates of black-hole spin; modeling the surrounding
environment of compact X-ray sources, such as a tenuous accretion-disk
corona \citep{White_1982} or a substantial corona that scatters
photons up to MeV energies \citep{Gierlinski_1999}; and classifying
patterns of distinct X-ray states, e.g., in black-hole binaries
(RM06).

Because of the importance of the power-law component, several physical
models have been developed to infer the conditions of the hot plasma
that causes the Comptonization.  Models of this variety that are
available in XSPEC are {\sc compTT} \citep{COMPTT}, {\sc eqpair} 
\citep{EQPAIR}, {\sc compTB}
\citep{COMPTB}, {\sc bmc} \citep{BMC}, {\sc compbb} \citep{COMPBB},
{\sc thcomp} \citep{THCOMP}, {\sc compls} \citep{COMPLS}, and {\sc compps}
\citep{COMPPS}.  It is
essential to use such physical models when one is focused on
understanding the physical conditions and structure of a scattering
corona or other Comptonizing plasma.

Often, however, the physical conditions of the Comptonizing medium are
poorly understood or are not of interest, and one is satisfied with an
empirical model that seeks to match the data with no pretense that the
model describes the physical system.  The model {\sc powerlaw} is one
such empirical model which has been extraordinarily widely used in
modeling black-hole and neutron-star binaries (see text \& references in
\citealt{White_1995,Tanaka_1995,Brenneman_2006}; RM06) and AGN
\citep[e.g.,][]{Zdziarski_2002,Brenneman_2006}.  However, {\sc
powerlaw} introduces a serious flaw: at low energies it rises without limit.
The divergence at low energies is unphysical, and it often
significantly corrupts the parameters returned by the model component
with which it is teamed (e.g., the widely used disk blackbody
component {\sc diskbb}; \S3).

An excellent alternative to the standard power-law model for describing
Compton scattering is provided by a convolution model that is based on a
Green's function that was formulated decades ago \citep{Shapiro_1976,Rybicki_Lightman,Sunyaev_1980,
COMPTT}.  In this approach the power-law is generated self-consistently
via Compton up-scattering of a seed photon distribution; consequently,
the power-law naturally truncates itself as the seed distribution falls
off at low energies.  

In this paper, we present our implementation of a flexible convolution
model named {\sc simpl} that can be used with any spectrum of seed
photons.  For a Planck distribution we show that {\sc simpl} gives 
identical results to {\sc bmc}, as expected since the two models are
functionally equivalent (\S2.3).  Although {\sc simpl} has only two
free parameters, the same number as the standard {\sc powerlaw}, this
empirical model is nevertheless able to very successfully fit data
simulated using {\sc compTT}, a prevalent physical model of
Comptonization (\S2.2).

We analyze data for two black hole binaries and illustrate the
flexibility of {\sc simpl} by convolving {\sc simpl} with {\sc diskbb},
the workhorse accretion disk model that has been used for decades
\citep{DISKBB}.  Our principal result is that {\sc simpl} in tandem with
{\sc diskbb} enables one to obtain fitted values for the inner-disk
radius $R_{\rm in}$ for strongly-Comptonized data that are
consistent with those obtained for weakly-Comptonized data (see \S3.2).
The standard power law, on the other hand, delivers very inconsistent
values of $R_{\rm in}$.  As we show in \S4.2, this result is
consequential for the measurement of black hole spin via the
continuum-fitting method: It implies that using {\sc simpl} in place of
the standard power law one can obtain reliable measurements of spin for
a far wider body of data than previously thought possible, and for more
sources (e.g., Cyg X-1).

In \S2 we outline the model and in \S3 we present a case study with
several examples.  We discuss the implications and intended applications
of the model in \S4 and conclude with a summary in \S5.


\section{The Model: {\sc simpl}}

The model {\sc simpl} (SIMple Power Law) functions as a convolution
that converts a fraction of input seed photons into a power law (see
eq. [\ref{eqn:convolution}]).  The model is currently available in
XSPEC\footnote{see
http://heasarc.nasa.gov/xanadu/xspec/manual/XSmodelSimpl.html}.  In
addition to {\sc simpl-{\small 2}}, which is our implementation of the classical
model described by \citet{Shapiro_1976} and \citet{Sunyaev_1980}, which 
corresponds to both up- and down-scattering of photons, we offer an 
alternative ``bare-bones'' implementation in which photons are only 
up-scattered in energy.  The physical motivations behind the two versions
 of the model are described in \S2.1, and the corresponding scattering 
kernels --- the Green's functions --- are given in equation (2) and 
equation (3), respectively.

The parameters of {\sc simpl} and the standard {\sc powerlaw} model
are similar.  Their principal parameter, the photon index $\Gamma$, is
identical.  However, in the case of {\sc simpl} the normalization
factor is the scattered fraction $\fsc$, rather than the photon flux.
The goal of {\sc simpl} is to characterize the effects of
Comptonization as simply and generally as possible.  In this spirit,
all details of the Comptonizing medium, such as its geometry (slab
vs. sphere) or physical characteristics (optical depth, temperature
,thermal vs. non-thermal electrons
), which would require additional parameters for their description, 
are omitted.

It is appropriate to employ {\sc simpl} when the physical conditions of
the Comptonizing medium are poorly understood or are not of interest.
When the details of the Comptonizing medium are known, or are the main
object of study, one should obviously use other models (e.g., {\sc
compTT}, {\sc compps}, {\sc thcomp}, etc.), which are designed
specifically for such work.  {\sc simpl}, on the other hand, is meant
for those situations in which a Compton power-law component is present
in the spectral data and needs to be included in the model but is not
the primary focus of interest.  {\sc simpl} should thus be viewed as a
broad-brush model with the same utility as {\sc powerlaw} but designed
specifically for situations involving Comptonization.

By virtue of being a convolution model, {\sc simpl} mimics physical
reprocessing by tying the power-law component directly to the energy
distribution of the input photons.  The most important feature of the
model is that it produces a power-law tail at energies larger than the
characteristic energy of the input photons, and that the power law does
not extend to lower energies.  This is precisely what one expects any
Compton-scattering model to do and is a general feature of all the
physical Comptonization models mentioned above.  In contrast, the model
{\sc powerlaw} simply adds to the spectrum a pure power-law component
that reaches all the way downward 
to arbitrarily low
energies.  The difference between {\sc simpl} and {\sc powerlaw} is thus
most obvious at soft X-ray bands where {\sc simpl} cuts off in a
physically natural way whereas {\sc powerlaw} continues to rise without
limit \citep[e.g., see][]{Yao_2005}.

Two assumptions underlie {\sc simpl}.  The first is that all soft
photons have the same probability of being scattered (e.g., the
Comptonizing electrons are distributed spatially uniformly).  This is
a reasonable assumption when one considers that, even in the best of
circumstances, almost nothing is known about the basic geometry of the
corona.  For example, usually the corona is variously and crudely
depicted as a sphere, a slab, or a lamp post.  The second assumption
is that the scattering itself is energy independent.  This is again
reasonable given the soft thermal spectra of the seed photons that are
observed for black-hole and neutron-star accretion disks with typical
temperatures of $\sim 1$ keV and a few keV, respectively.  For
example, in the extreme case of a $180^\circ$ back-scatter off a
stationary electron, a 3 keV seed photon suffers only a 1\% loss of
energy, and even a 10 keV photon loses only 4\% of its initial energy.

Figure~1 shows sample outputs from {\sc simpl} when the input soft
photons are modeled by the multi-temperature disk blackbody model {\sc
diskbb} \citep{DISKBB}.  Results are shown for both {\sc simpl-{\small 2}} and
{\sc simpl-{\small 1}}, our alternative version of {\sc simpl} that
includes only up-scattering of photons; the spectra are shown for
$\Gamma=2.5$ and a range of values of $\fsc$.  Note the power-law
tails in the model spectra at energies above the peak of the soft
thermal input and the absence of an equivalent power-law component at
lower energies.  This is the primary distinction between {\sc simpl}
and {\sc powerlaw}.  {\sc simpl-{\small 2}} and {\sc simpl-{\small 1}} give
similar spectra, but the spectrum from {\sc simpl-{\small 1}} has a
somewhat stronger power-law tail for the same value of $\fsc$.  This
is because {\sc simpl-{\small 1}} transfers all the scattered photons
to the high energy tail, whereas {\sc simpl-{\small 2}} has double-sided
scattering.  Therefore, for the same value of $\fsc$, fewer photons
are scattered into the high-energy tail with {\sc simpl-{\small 2}}.
Correspondingly, when fitting the same data, {\sc simpl-{\small 2}} returns a
larger value of $\fsc$ compared to {\sc simpl-{\small 1}} (for
examples, see \S3 and Table~2).

\subsection{Green's Functions}

Given an input distribution of photons $\nin (E_0)dE_0$ as a function of
photon energy $E_0$, {\sc simpl} computes the output distribution $\nout
(E)dE$ via the integral transform:
\begin{equation}
\nout(E)dE = (1-\fsc)\nin(E)dE+\fsc
\left[\int_{E_{\rm min}}^{E_{\rm max}} \nin(E_0) G(E;E_0)dE_0\right] dE.
\label{eqn:convolution}
\end{equation}
A fraction $(1-\fsc)$ of the input photons remains unscattered (the
first term on the right), and a fraction $\fsc$ is scattered (the
second term).  Here, $E_{\rm min}$ and $E_{\rm max}$ are the minimum
and maximum photon energies present in the input distribution, and
$G(E;E_0)$ is the energy distribution of scattered photons for a
$\delta$-function input at energy $E_0$, i.e., $G(E;E_0)$ is the
Green's function describing the scattering.

We now describe the specific prescriptions we use for {\sc simpl-{\small 2}} and
{\sc simpl-{\small 1}}.  We also discuss the physical motivations behind
these prescriptions, drawing heavily on the theory of Comptonization as
described by \citet[][hereafter RL79]{Rybicki_Lightman}.

\subsubsection{{\sc simpl-{\small 2}}}

In sec.~7.7, RL79 discuss the case of unsaturated repeated scattering
by nonrelativistic thermal electrons.  Following \citet*{Shapiro_1976},
they solve the Kompaneets equation and show that
Comptonization produces a power-law distribution of photon energies
(eq. 7.76d in RL79).  There are two solutions for the photon index
$\Gamma$:
\begin{eqnarray*}
\Gamma_1 &=& -{1\over2} + \sqrt{{9\over 4}+{4\over y}}, \\
\Gamma_2 &=& -{1\over2} - \sqrt{{9\over 4}+{4\over y}},
\end{eqnarray*}
where the Compton $y$ parameter is given by $y=(4kT_e /m_e c^2){\rm
Max}(\tau_{\rm es}, \tau_{\rm es}^2)$.  Up-scattered photons have a
power-law energy distribution with photon index $\Gamma_1$ and
down-scattered photons have a different power-law distribution with
photon index $\Gamma_2$.

We model this case of nonrelativistic electrons with the following
Green's function \citep{Sunyaev_1980, COMPTT,Ebisawa_99}, 
 which corresponds to the model {\sc simpl-{\small 2}}:
\begin{eqnarray}\label{eqn:simpl2}
G(E;E_0)dE=\frac{(\Gamma-1)(\Gamma+2)}{(1+2\Gamma)} \begin{cases}
   (E/E_0)^{-\Gamma} dE/E_0, \;\qquad E\geq E_{0} \\
   (E/E_0)^{\Gamma+1} dE/E_0,\qquad E < E_0. 
\end{cases}
\end{eqnarray}
The function is continuous at $E=E_0$, is normalized such that it
conserves photons, and holds for all $\Gamma>1$.  Substituting
(\ref{eqn:simpl2}) in (\ref{eqn:convolution}) we see that {\sc simpl-{\small 2}} has
two parameters: $\fsc$ and $\Gamma$.  Although the model makes use of
two power laws, their slopes are not independent.

As in the case of the standard power law, {\sc simpl} includes no high
energy cutoff.  Technically, for any complete model of Comptonization,
the up-scattered power-law distribution is cut off for photon energies
larger than $kT_e$.  To avoid increasing the complexity of our model,
we have ignored this detail; extra parameters could easily be added to
account for high energy attenuation if desired.  By keeping the model
very basic, {\sc simpl} is a direct two-parameter replacement for the
standard power law while bridging the divide between the latter model
and physical Comptonization models.

\subsubsection{{\sc simpl-{\small 1}}}

The Green's function (2) is obtained by solving the Kompaneets
equation, which assumes that the change in energy of a photon in a
single scattering is small.  This assumption is not valid when the
Comptonizing electrons are relativistic.

In sec.~7.3 of their text, RL79 discuss Compton scattering by
relativistic electrons with a power-law distribution of energy:
$n_e(E_e)dE_e \propto E_e ^{-p}dE_e$.  In the limit when the optical
depth is low enough that we only need to consider single scattering,
they show that the Comptonized spectral energy distribution (SED) is a
power law of the form $P(E)dE \propto E^{-(p-1)/2}$.  Equivalently,
the photon energy distribution takes the form $n(E)dE \propto
E^{-\Gamma}$, with a photon index $\Gamma = (p+1)/2$.  Hardly any
photons are down-scattered in energy.

In sec.~7.5, RL79 show that repeated scatterings produce a power-law SED
even when the relativistic electrons have a non-power-law distribution
\citep[see also][]{Titarchuk_95}.  In terms of the mean amplification of
photon energy per scattering $A$ and the optical depth to electron
scattering $\tau_{\rm es}$, the Comptonized photon energy distribution
takes the form $n(E)dE \propto E^{-\Gamma}$ with a photon index $\Gamma
= 1-\ln\tau_{\rm es}/\ln A$.  For the specific case of a thermal
distribution of electrons with a relativistic temperature $kT_e \gg
m_ec^2$, the amplification factor is given by $A=16(kT_e /m_e c^2)^2$.
Once again, hardly any photons are down-scattered.

For both cases discussed above, Comptonization is dominated by
up-scattering and produces a nearly one-sided power-law distribution
of photon energies.  This motivates the following Green's function,
valid for $\Gamma>1$, which we refer to as the model {\sc
simpl-{\small 1}}:
\begin{eqnarray}\label{eqn:simpl1}
G(E;E_0)dE = \begin {cases}
  (\Gamma-1)({E/E_0})^{-\Gamma}dE/E_0, ~~E \geq E_0 \\
  0, \qquad\qquad\qquad\qquad\quad  E < E_0. 
\end{cases}
\end{eqnarray}
The normalization factor $(\Gamma-1)$ ensures that we conserve
photons.  

Although {\sc simpl-{\small 1}} is most relevant for relativistic
Comptonization, it can also be used as a stripped-down version of
{\sc simpl-{\small 2}} for non-relativistic coronae.  The reason is that the
low-energy power-law $(E/E_0)^{\Gamma+1}$ in equation (2) almost never
has an important role.  There is not much power in this component, and
what little contribution it makes is indistinguishable from the input
soft spectrum.  Therefore, even for the case of nonrelativistic
thermal Comptonization, for which the Green's function (2) is
designed, there would be little difference if one were to use {\sc
simpl-{\small 1}} instead of {\sc simpl-{\small 2}}.

\subsection{Comparison to {\sc compTT}}

To illustrate the performance of {\sc simpl} relative to other
Comptonization models, we have simulated a $2 \times 10^6$-count {\it BeppoSAX}
\citep{BEPPOSAX} observation using the {\sc compTT} model in XSPEC
v12.4.0x.

For our source spectrum, we adopt disk geometry, a Wien distribution of
seed photons at $kT_0 = 1 \; \keV$, and a hydrogen column density of
$\nh = 10^{21} \; \cm ^{-2}$.  We set the optical depth and
temperature of the Comptonizing medium to $\tau_c = 2$ and $kT_e = 40
\; \keV$.  Our simulation uses the LECS, MECS, and PDS detectors on {\it
BeppoSAX}, which span a wide energy range $\sim 0.1-200$ keV (for
details on the instruments, see \S3). The total
number of counts in the simulated spectra ($\sim 2 \times 10^6$) corresponds to a
3 ks observation of a 1 Crab source.

We analyze the simulated data with a model consisting of a blackbody
({\sc bb}) coupled with {\sc simpl}.  We refer to this model
as {\sc simpl$\otimes$bb} (the {\sc$\otimes$} is to emphasize that
{\sc simpl} represents a convolution).  The best fits achieved have
reduced chi-squared values of $\chi_\nu^2=1.00$ ({\sc simpl-{\small 1}}) 
and $\chi_\nu^2=1.06$ ({\sc simpl-{\small 2}}).  The
fitted {\sc bb} temperatures are respectively $1.14 \pm 0.02 $ keV and
$1.29 \pm 0.01$ keV compared to 1 keV in the original {\sc compTT}
model.   Figure~2 shows the fit using {\sc simpl-{\small 1}} 
and Table~1 lists the best-fit parameters for both models.

In comparison, {\sc compbb}, an alternative model of Compton
scattering that assumes slab geometry, fits our simulated spectrum 
comparably well as 
 {\sc simpl}, with $\rchi = 1.05$ (Table~1).  {\sc compbb}
returns the same temperature as {\sc simpl-{\small 2}}, $kT_{\rm
bb}=1.29 \pm 0.01$ keV.  Compared to the {\sc compTT} progenitor, {\sc
compbb} gives similar estimates of the coronal temperature
$kT_e$ and optical depth $\tau_c$ (Table~1).  Even though {\sc
compbb} is a physically more realistic model of coronal scattering
than {\sc simpl}, it does not outperform {\sc simpl} in terms of
fitting the {\sc compTT}-generated data.  Meanwhile, the model 
{\sc bb}+{\sc powerlaw} performs quite poorly, yielding $\rchi > 2$.  
Parameters for this fit
are given in Table~1.  Note that the derived $\nh$ using {\sc
powerlaw} is much higher than either the original value or those from
fits with {\sc simpl}.

Though {\sc simpl} is a purely empirical model, we see that it can
deliver a remarkably successfully fit to data simulated using the
physical model {\sc compTT}.  Even for a very cool corona with electron
temperatures as low as $kT_e = 20$ keV, which causes {\sc compTT} to
produce noticeable curvature in the high-energy spectrum, we find that
{\sc simpl-{\small 2}} and {\sc simpl-{\small 1}} achieve reasonable fits with
$\rchi < 1.2$.

A significant virtue of {\sc simpl} relative to the physical
Comptonization models in XSPEC is that {\sc simpl} can be employed in
conjunction with any source of seed photons.  The physical models, on
the other hand, are typically restricted to treating only one or two
predefined photon distributions.  One standard choice of continuum
model that is widely used in fitting Comptonized accretion disks is
{\sc diskbb}+{\sc compTT}.  With {\sc simpl}, one would instead employ
the model {\sc simpl$\otimes$diskbb}.  The latter not only generates
the power law self-consistently via up-scattering of the seed photons,
but it also has two fewer parameters.

\subsection{Bulk Motion Comptonization}

The model {\sc bmc} describes the Comptonization of blackbody seed
photons by a converging flow of isothermal gas that is freely falling
toward a compact object, i.e., bulk motion Comptonization \citep[see,
e.g., ][]{Shrader_1998,BMC}.  {\sc bmc} is an alternative to coronal
Comptonization models and is structured {\it identically} to {\sc
simpl-{\small 2}$\otimes$bb}; both models are specified with just four parameters.
As a direct demonstration in XSPEC that {\sc simpl-{\small 2}$\otimes$bb} and {\sc
bmc} are identical, we analyzed our simulated {\it BeppoSAX} spectrum
described above using both models.  We found that the returned values of
the column density $N_{\rm H}$, the blackbody temperature $kT$, and the
photon index $\Gamma$ agreed in each case to four or more significant
figures.

{\sc bmc} has been variously used to support claims that Compton
scattering off in-falling gas within several gravitational radii gives
rise to the observed high energy power law in several black-hole
binaries \citep[e.g.,][]{Shrader_1998,Shrader_1999,Borozdin_1999}.
However, this is only one interpretation of the model; {\sc
simpl-{\small 2}}$\otimes${\sc bb}, and therefore {\sc bmc}, can equally
be used to support a more standard model of coronal scattering
(operating with uniform efficiency at all energies, see \S2 and
\S2.1.2).  Thus, although {\sc bmc} is designed specifically to model
relativistic accretion inflows, its function is actually quite
general.

One virtue of {\sc simpl} relative to {\sc bmc} is that {\sc simpl}
does depend upon discerning 
the nature of the Comptonizing region, be
it corona, relativistic in-falling gas, or other.  Another virtue of
{\sc simpl} is that it fully incorporates the utility of {\sc bmc}
while allowing complete flexibility in the choice of the spectrum of
seed photons, e.g., {\sc simpl$\otimes$diskbb} is more appropriate for
modeling Comptonization in accretion disks than {\sc bmc}, which is
hardwired to a Planck function.

The theory of bulk motion Comptonization is developed further and
rigorously in \citet{BMC}.  This paper describes a Green's function
that is more appropriate than the one used in {\sc bmc}.  A complete
version of this Green's function is incorporated into the more
sophisticated model {\sc compTB}.  However, this model is again
limited to treating scattering from a predefined set of
(blackbody-like) seed photon distributions and includes additional
free parameters.  We find that the fitting results obtained using this
Green's function are intermediate between those given by {\sc simpl-{\small 1}}
and {\sc simpl-{\small 2}} so long as the temperature of the
in-flowing electrons, $T_e$, is above the observed energy range.


\section{Data Analysis}

In this section, we apply {\sc simpl} to a sample of observations to
illustrate how {\sc simpl} compares with {\sc powerlaw}.  To this end,
we have selected two black-hole binaries, H1743--322 and LMC~X--3.
H1743--322 (hereafter H1743) is an especially pristine black-hole
transient \citep[see ][]{RR_H1743_2006} since, for much of its 2003
outburst, its spectrum can be satisfactorily modeled with just absorbed
($\nh \approx 2.2\times10^{22} \cm^{-2}$) thermal-disk and power-law
components \citep[][hereafter M07]{JEM_H1743_2007}.  In particular, the
122 days of contiguous spectral data on which we focus do not require
any additional components to accommodate the reflection or absorption
features that are often present in the spectra of black hole binaries.

The spectra of H1743 were acquired by the {\it Rossi X-ray Timing
Explorer} ({\it RXTE}) PCU-2 module \citep{RXTE}, {\it RXTE}'s
best-calibrated PCU detector, and were taken in ``standard 2'' format.
All spectra have been background subtracted and have typical exposure
times $\sim3000\;$s.  The customary systematic error of 1\% has been
added to all energy channels.  
The resultant pulse-height spectra
are analyzed from $2.8 - 25$ keV using XSPEC v12.4.0x (see M07 for
further details).

While {\it RXTE} provides good spectral coverage in hard X-rays
($\gtrsim 10 \;\kev$), which is most important for constraining the
power-law component, it is not sensitive at low energies ($<2.5
\;\kev$).  Therefore, {\it RXTE} data are generally insensitive to
$\nh$.  To complement the {\it RXTE} observations presented here, we
have selected a {\it BeppoSAX} observation of LMC X--3, a persistent and
predominantly thermal black-hole source with a very low hydrogen column
\citep[$\nh \approx 4\times10^{20} \cm^{-2}$;][]{Page_2003, Yao_2005}.

The {\it BeppoSAX} narrow-field instruments provide sensitive
measurements spanning a wide range in energy, from tenths to hundreds of
keV.  The low-energy concentrator system (LECS) and the medium-energy
concentrator system (MECS) probe soft fluxes, from $\sim0.1-4$ keV and
$\sim1.5-10$ keV, respectively.  The phoswich detector system (PDS) is
sensitive to hard X-rays from $\sim15-200$ keV, and the high-pressure
gas scintillation counter (HPGSPC) covers $\sim4-100$ keV.  In this
analysis, we consider only the LECS, MECS, and PDS because the
statistical quality of the HPGSPC data is relatively poor.

In reducing {\it BeppoSAX} data, we have followed the protocols given in
the Cookbook for {\it BeppoSAX} NFI Spectral Analysis
\citep{BepposaxABC}.  We use pipeline products and extract spectra from
8$\arcmin$ apertures centered on LMC X--3 for both the LECS and
(combined) MECS detectors.  For the PDS, which is a simple collimated
phoswich detector, we selected the fixed rise-time spectrum.  In our
analysis, we have used standard response matrices and included
blank-field background spectra with the appropriate scalings.  No
pile-up correction is necessary.

\subsection{Steep Power Law State}

About a third of the way through its nine-month outburst cycle, H1743
repeatedly displayed spectra in the steep power-law (SPL) state that were 
devoid of  absorption features.  A salient feature of the SPL state is the 
presence of a strong power-law component of emission.  (For a review of
black-hole spectral states and a precise definition of the SPL state,
see Table~2 and text in RM06.) Twenty-eight such featureless spectra
were consecutively observed over a period of about three weeks (spectra
\#58--85; M07).  We focus here on one representative spectrum, \#77.  In
Figure~3 we show our fits and the associated unabsorbed models obtained
using {\sc diskbb}+{\sc powerlaw} and {\sc simpl$\otimes$diskbb}.
Fitted spectral parameters are presented in Table~2.

The quality of fit (as measured by $\rchi$) using either model is
comparable.  Nevertheless, there are distinct differences between the
models.  The fits with {\sc simpl} have a $\sim 50$\% larger disk
normalization compared to {\sc powerlaw} and a $\sim 40$\% lower $\nh$
(Table~2).  The fit using {\sc powerlaw} diverges at low energies, as
revealed by removing photoabsorption from the fitted models (panels on
the right in Fig. 3).  The effect is quite severe and has no obvious
physical explanation.  In contrast, the fit using {\sc simpl} is well
behaved and the unabsorbed model is not divergent.

\subsection{Thermal Dominant State}

The key feature of the thermal dominant (TD) state is the presence of a totally dominant
and soft ($kT \sim 1$ keV) blackbody-like component of emission that
arises in the innermost region of the accretion disk.  The TD state is
defined by three criteria, the most relevant of which here is that the
fraction of the total 2--20 keV unabsorbed flux in the thermal component
is $\ge 75$\%.  For the full definition of this state, see Table~2 in
RM06.

Here we have chosen H1743 spectrum \#91 which belongs to a sequence of
$\sim$50 featureless spectra (\#86--136; M07) in the TD state.  
This spectrum has $\Gamma \sim 2$, which is somewhat harder
than usual, but is otherwise typical of H1743's TD state.  Spectral fit
results are shown in Figure~4.  In addition, in order to further
illustrate for the TD state the differences between {\sc simpl} and {\sc
powerlaw} at energies below the $\approx 2.5$ keV response cutoff of
{\it RXTE}, we use a {\it BeppoSAX} observation of LMC X--3; our results
are illustrated in Figure~5.  This observation was carried out on 1996
November 28 with exposure times of 1.8, 4.5, and 2 ks respectively for
the LECS, MECS and PDS.

As in \S3.1, we fit these data using {\sc diskbb+powerlaw}
and {\sc simpl$\otimes$diskbb}.  The best-fit spectral parameters are
listed in Table~2.  Due to a calibration offset between
the various {\it BeppoSAX} instruments, we follow standard procedure and
fit for the normalization of the LECS and PDS relative to the MECS, the
best-calibrated of the three.  We adopt the canonical limits of $0.7-1$
for LECS/MECS and $0.77-0.93$ for PDS/MECS.  These normalizations are
included in the tabulated results.

A comparison of the results obtained with {\sc powerlaw} and {\sc simpl}
confirms the trends highlighted in \S3.1, namely the
differences in normalization and $\nh$.  However, they are more modest
here because the Compton component is weaker in the TD state.

\subsection{Comparison of {\sc simpl} and {\sc powerlaw}}

An examination of Table~2 reveals the following systematic
differences in the derived spectral parameters returned when fitting
with {\sc simpl} vs. {\sc powerlaw}: {\sc simpl} yields (i) a stronger
and softer thermal disk component, i.e., a larger normalization and
lower $kT_*$; (ii) a generally steeper power law component (larger
$\Gamma $); and (iii) a systematically lower $\nh$.  As we now show, all
of these effects can be simply understood.

Because {\sc powerlaw} produces higher fluxes than {\sc simpl} at low
energies, it tends to suppress the flux available to the (soft)
thermal component, namely {\sc diskbb} in the examples given here.
This explains why {\sc powerlaw} tends to harden the {\sc diskbb}
component and to steal flux from it (i.e., reduce its normalization
constant).  Meanwhile, at low energies the {\sc powerlaw} component
predicts artificially high fluxes that, in order to conform to the
observed spectrum, depress the value of $\Gamma$.  These differences
between {\sc simpl} and {\sc powerlaw} are most pronounced when the
power law is relatively steep, i.e., typically when $\Gamma \gtrsim
3$.

Modest and reasonable values of $\nh$ are returned in fits using {\sc
simpl}, as well as {\sc compTT} and other Comptonization models, because
the Compton tail is produced by the up-scattering of seed photons and
there is no power-law component at low energies.  In contrast, {\sc
powerlaw} continues to rise at low energies, which forces $\nh$ to
increase in order to allow the model to fit the observed spectrum.  This
systematic difference is apparent in our fit results for the H1743
spectra and is especially prominent in the case of the LMC X-3 spectrum
for which $\nh$ differs by a factor of two.  For H1743, the discrepancy
in $\nh$ is much less for the TD spectrum than for the SPL spectrum
because the SPL state has both a steeper and relatively stronger
power-law component.

We turn now to consider the {\sc diskbb} normalization constant, which
is proportional to $R_{\rm in}^2$, the square of the inner disk radius
(see footnotes to Table~2).  For the pair of H1743 spectra, we see that
the disk normalization obtained with {\sc powerlaw} is $\approx$35\%
smaller in the SPL state than in the TD state (Table~2), indicating that
$R_{\rm in}$ is smaller for the SPL state.  With {\sc simpl}, on the
other hand, there is no significant change in the normalization, and
hence both the SPL and TD states can be modeled with a disk that has the
same inner radius.  The radius is constant because {\sc simpl} recovers
the original (unscattered) flux emitted by the disk, which {\sc
powerlaw} cannot do.

This crucial ability to unify the inner regions of the accretion disk 
in thermally active states exhibiting both high and low levels of Comptonization 
(i.e., TD and SPL states) paves the way for a full general relativistic analysis which 
can formally link $\rin$ to black-hole spin (see discussion in \S\ref{subsec:spin}). 
\citet{Kubota_2004} similarly identified a constant radius for the black hole binary
XTE~J1550--564 between the TD and SPL states in an analysis using the
model {\sc diskbb + thcomp}.  Because {\sc thcomp} is implemented as an 
additive (i.e., non-convolution) model, Kubota \& Makishima had to employ an awkward and
ad hoc procedure to obtain their result (see their Appendix).  Their work improved upon a similar result obtained for black-hole GRO~J1655--40 \citep{Kubota_2001}.  
With {\sc
simpl}, the modeling is significantly easier.


\section{Discussion}

\subsection{Black Hole X-ray States}

A standard method of classifying X-ray states in black hole binaries
involves spectral decomposition into two primary components -- a
multi-temperature blackbody disk, {\sc diskbb}, and a Compton power law,
{\sc powerlaw} (RM06).  This method is compromised by the use of the
standard power law when the photon index is large ($\Gamma \gtrsim 3$).
In this case, at low energies the flux from the power law can rival or
exceed the thermal component and thereby pollute it.  As discussed in
\S2, intrusion of the power-law component at low
energies is fundamentally inconsistent with Compton scattering.

This difficulty in classifying states, which is caused by the use of
{\sc powerlaw}, is remedied by the use of {\sc simpl} because the latter
model naturally truncates the power-law component at low energies.  It
is useful to consider the intrinsic differences between the two models
and how they influence the classification of black-hole X-ray states.
Using {\sc powerlaw}, the thermal disk and tandem power-law emission are
modeled independently.  On the other hand, under {\sc simpl} all
photons originate in the accretion disk.  Some of these disk photons
scatter into a power law en~route from the disk to the observer.   As
described in \S3.3, fits employing {\sc simpl} imply stronger disk
emission and weaker Compton emission than those using {\sc powerlaw}.
As a result, state selection criteria would need to be modified for
classification using {\sc simpl}. This topic is beyond the scope of the
present paper.

\subsection{Application to the Measurement of Black Hole Spin}\label{subsec:spin}

During the past few decades, the masses of 22 stellar black holes have
been measured, 17 of which are found in transient black hole binaries
(RM06).  Recently, we have measured the spins of four of these stellar
black holes \citep{Shafee_spin, spin_1915, spin_m33} by fitting their
continuum spectra to our fully relativistic model of an accretion disk
{\sc kerrbb2} \citep{KERRBB, spin_1915} plus the standard power-law
component {\sc powerlaw}.  
In the continuum-fitting method, spin is measured
by estimating the inner radius of the accretion disk $R_{\rm in}$ 
\citep{Zhang_1997}.  We
identify $R_{\rm in}$ with the radius of the innermost stable circular
orbit $R_{\rm ISCO}$, which is predicted by general relativity.  

To date, we have conservatively selected only
TD data for analysis (\S3.2). 
Meanwhile, the transient black
hole binaries spend only a modest fraction of their outburst cycle in
the TD state and are often found in the SPL or some intermediate state
(see Figs. 4--9 in RM06).  Since for each source we seek to obtain as
many independent measurements of the spin parameter as possible, our
sole reliance on TD data has been a significant limitation.  
Using {\sc simpl}, the highly Comptonized SPL state is now able to provide
estimates of spin, a matter to be discussed more fully in 
(J. Steiner et al.\ 2009, in preparation). 
Not only will this allow us to substantially
increase the size of our data sample for many sources, it also will
likely allow us to obtain spin measurements for sources such as Cygnus
X-1 that never enter the TD state \citep{MR06}.


Our reason for developing {\sc simpl} was to improve our methods for
analyzing TD-state data in order to determine more reliable values of
black hole spin.  That {\sc simpl} now allows us to determine spins
for SPL-state data was a serendipitous discovery and a major
bonus.  We were motivated to develop {\sc simpl} because we have been
hampered by the use of {\sc powerlaw} in two ways.  First, in all of
our work we have exclusively used TD spectral data
\citep[e.g.,][]{Shafee_spin}, which is maximally free of the uncertain
effects of Comptonization.  The selection of these data is problematic
because, as indicated in \S4.1 above, it can be affected in unknown
ways when the spectral index of the Compton component is large.

Secondly, of greater concern is the potential adulteration in the TD
state of the thermal component by the power-law component, which can
have an uncertain and sizable effect on the fitted value of the
black-hole spin parameter.  In the case of a number of
spectra with steep power-law components, we found that the fitted values
of the spin parameter were affected by the contribution of the power law
flux at energies below $\sim 5$ keV; e.g., see \S~4.2 in
\cite{spin_1915}.  In order to mitigate this problem, we applied in turn
two alternative models: {\sc compTT}, and a standard power-law component
curtailed by an exponential low-energy cutoff, {\sc
expabs$\times$powerlaw}.  The former model was unsatisfactory because we
were unable to fit for reasonable values of both the coronal temperature
$kT_e$ and the optical depth $\tau_c$.  The latter model was likewise
unsatisfactory because it requires the use of an arbitrary cutoff energy
$E_c$.

We developed {\sc simpl} in order to sidestep these difficulties and
uncertainties.  At low energies, the model truncates the power law in
the same physical manner as {\sc compTT} and other sophisticated
Comptonization models.  {\sc simpl} self-consistently ties the
emergent power-law flux to the seed photons in order to deliver the
power-law component via coronal reprocessing.  An application of {\sc
simpl} to the measurement of the spin of the black-hole primary in LMC
X-1 is described in L.\ Gou et al. (2009).


\section{Summary}

We present a new prescription for treating Comptonization in X-ray
binaries.  While no new physics has been introduced by this model, its
virtues lie in its simplicity and natural application to a wide range of
neutron-star and black-hole X-ray spectra.  {\sc simpl} offers a generic
and empirical approach to fitting Comptonized spectra using the minimum
number of parameters possible (a normalization and a slope), and it is
valid for a broad range of geometric configurations (e.g.,
uniform slab and spherical geometries).  The scattering of a seed
spectrum occurs via convolution, which self-consistently mimics physical reprocessing of
photons from, e.g., an accretion disk.  In addition to this physically
motivated underpinning, {\sc simpl} remains as unassuming as {\sc
powerlaw} but without its troublesome divergence at low energies.

Our model is valid for all $\Gamma>1$.  We have shown that {\sc simpl}
is able to provide a good fit to a demanding simulated data set, which
was generated with the widely-used Comptonization model {\sc compTT}.
Furthermore, we have demonstrated that {\sc simpl} and {\sc powerlaw}
give very comparable results when fitting spectral data in terms of
quality-of-fit and spectral index (see Table \ref{tab:H1743}).  This
quality of performance holds true not only for spectra with weak
Compton tails (TD state) but also for spectra requiring a large
Compton component (SPL state).  In the latter case, the model based on
{\sc simpl} gives physically more reasonable results for the soft end
of the spectrum (e.g., see \S3.3).

Using {\sc simpl$\otimes$diskbb} it will be important to revisit the
classification of black hole states (RM06) for two reasons.  First, the
selection of TD data will no longer be adversely affected by the
presence of a steep power-law component.  Secondly, this model will
allow some degree of unification of the TD state and SPL state, the
latter being a more strongly Comptonized version of the former.  In
determining black hole spin via the continuum-fitting method using {\sc
kerrbb2}, {\sc simpl} is a significant advance on three fronts: It will
(1) enable the selection of data with a dominant thermal component that
is not mucked up by the effects of a divergent power-law component; (2)
allow reliable spin measurements to be obtained using
strongly-Comptonized SPL data, thereby substantially increasing the data
sample for a given source; and (3) likely make possible the
determination of the spins of some black holes that do not enter the TD
state (e.g., Cygnus X-1).


\acknowledgements The authors would like to thank George Rybicki for
discussions on the physics of Comptonization as well as Jifeng Liu,
Lijun Gou, Rebecca Shafee, and Ron Remillard for their input on {\sc
simpl}.  JFS thanks Joey Neilsen for enthusiastic discussions as well as
comments on the manuscript, Ryan Hickox for suggestions which improved
this paper, and Keith Arnaud for helping implement {\sc simpl} in XSPEC.
The authors thank Tim Oosterbroek for his indefatigable assistance with
the {\it BeppoSAX} reduction software.  JFS was supported by the
Smithsonian Institution Endowment Funds and RN acknowledges support from
NASA grant NNX08AH32G and NSF grant AST-0805832.  JEM acknowledges
support from NASA grant NNX08AJ55G.


\appendix
\section{XSPEC Implementation}

{\sc simpl} is presently implemented in XSPEC.  This version includes
three parameters (two that can be fitted), the power-law photon index
($\Gamma$), the scattered fraction ($\fsc$), and a switch to set 
up-scattering only ({\sc simpl-{\small 1}}: switch~$>0$) and
double-sided scattering ({\sc simpl-{\small 2}}: switch~$\leq 0$). 
Since {\sc simpl} redistributes input photons to higher (and lower) energies, 
for detectors with limited response matrices (at high or low energies), or
poor resolution, the sampled energies should be extended or resampled
within XSPEC to adequately cover the relevant range.  For example,
when treating the {\it RXTE} data in \S3, which has no response
defined below $1.5$ keV, the command ``{energies 0.05 50 1000 log}''
was used to explicitly extend and compute the model over 1000
logarithmically spaced energy bins from $0.05-50$ keV.

Using {\sc simpl} can be problematic when $\Gamma$ is large, especially
if the power-law component is faint or the detector response extends
only to $\sim 10$ keV (e.g., {\it Chandra}, {\it XMM} or {\it ASCA}).
When the photon index becomes sufficiently large, a runaway process can
occur in which $\Gamma$ steepens and the scattered fraction becomes
abnormally high (typically $\gtrsim 50\%$, inconsistent with a weak
power law).  This occurs because scattering redirects photons from
essentially a $\delta$-function into a new function with characteristic
width set by $\Gamma$.  If $\Gamma$ reaches large values ($\gtrsim 5$),
the scattering kernel will also act like a $\delta$-function, and the
convolved spectrum will be nearly identical to the seed spectrum.

In such circumstances, we recommend bracketing $\Gamma$.  In practice,
the power-law spectral indices of black-hole systems are found to lie in
the range $1.4 \lesssim \Gamma \lesssim 4$ \citep{RR_JEM_review_2006}.
An upper limit of $\Gamma \sim4-4.5$ is typically sufficient to prevent
this runaway effect, and this constraint should be applied if it is
deemed appropriate for the source in question.

We advise against applying {\sc simpl} to sharp components such as
spectral lines or reflection components.  {\sc simpl} broadens spectral
lines, and for prominent higher energy emission features (such as Fe
K$\alpha$), {\sc simpl} can saturate the power-law flux with scattering
from the line itself.  To prevent this from occurring, such components should be invoked
outside the scope of {\sc simpl}.  For example, the XSPEC model
declaration ``{model phabs$\times$(smedge$\times$simpl(kerrbb)+laor)}''
would satisfy this recommendation.


\newcounter{BIBcounter} 
\refstepcounter{BIBcounter} \bibliography{ms} \mbox{~}

\begin{thebibliography}{41}
\expandafter\ifx\csname natexlab\endcsname\relax\def\natexlab#1{#1}\fi

\bibitem[{{Arnaud}(1996)}]{XSPEC} {Arnaud}, K.~A. 1996, in ASP Conf.\
Series, Vol. 101, Astronomical Data Analysis Software and Systems V,
ed. G.~H.  {Jacoby} \& J.~{Barnes} (San Francisco: ASP), 17

\bibitem[{{Boella} {et~al.}(1997){Boella}, {Butler}, {Perola}, {Piro},
  {Scarsi}, \& {Bleeker}}]{BEPPOSAX}
{Boella}, G., {Butler}, R.~C., {Perola}, G.~C., {Piro}, L., {Scarsi}, L., \&
  {Bleeker}, J.~A.~M. 1997, \aaps, 122, 299

\bibitem[{{Borozdin} {et~al.}(1999){Borozdin}, {Revnivtsev}, {Trudolyubov},
  {Shrader}, \& {Titarchuk}}]{Borozdin_1999}
{Borozdin}, K., {Revnivtsev}, M., {Trudolyubov}, S., {Shrader}, C., \&
  {Titarchuk}, L. 1999, \apj, 517, 367

\bibitem[Brenneman \& Reynolds(2006)]{Brenneman_2006}
Brenneman, L. W., \& Reynolds, C. S. 2006, \apj, 652, 1028

\bibitem[{{Cowley} {et~al.}(1983){Cowley}, {Crampton}, {Hutchings},
  {Remillard}, \& {Penfold}}]{Cowley_1983}
{Cowley}, A.~P., {Crampton}, D., {Hutchings}, J.~B., {Remillard}, R., \&
  {Penfold}, J.~E. 1983, \apj, 272, 118

\bibitem[Coppi(1999)]{EQPAIR} {Coppi}, P.~S.\ 1999, High Energy 
Processes in Accreting Black Holes, ed. J.~{Poutanen} \& R.~{Svensson} (San Francisco: ASP), 161, 375 

\bibitem[{{Ebisawa}(1999)}]{Ebisawa_99} {Ebisawa}, K. 1999, in ASP
Conf.\ Series, Vol. 161, High Energy Processes in Accreting Black Holes,
ed. J.~{Poutanen} \& R.~{Svensson} (San Francisco: ASP), 39

\bibitem[{{Farinelli} {et~al.}(2008){Farinelli}, {Titarchuk}, {Paizis}, \&
  {Frontera}}]{COMPTB}
{Farinelli}, R., {Titarchuk}, L., {Paizis}, A., \& {Frontera}, F. 2008, \apj,
  680, 602

\bibitem[{{Fiore} {et~al.}(1999){Fiore}, {Guainazzi}, \&
  {Grandi}}]{BepposaxABC}
{Fiore}, F., {Guainazzi}, M., \& {Grandi}, P. 1999, {Cookbook for BeppoSAX NFI
  Spectral Analysis}, Bepposax SDC


\bibitem[{{Gierli{\'n}ski} {et~al.}(1999){Gierli{\'n}ski}, {Zdziarski},
  {Poutanen}, {Coppi}, {Ebisawa}, \& {Johnson}}]{Gierlinski_1999}
{Gierli{\'n}ski}, M., {Zdziarski}, A.~A., {Poutanen}, J., {Coppi}, P.~S.,
  {Ebisawa}, K., \& {Johnson}, W.~N. 1999, \mnras, 309, 496



\bibitem[Gou et al.(2009)]{Gou} Gou, L., McClintock, J.~E., 
Liu, J., Narayan, R., Steiner, J.~F., Remillard, R.~A., Orosz, J.~A., 
\& Davis, S.~W.\ 2009, arXiv:0901.0920 [astro-ph]



\bibitem[Kubota et al.(2001)]{Kubota_2001} Kubota, A., Makishima, 
K., \& Ebisawa, K.\ 2001, \apjl, 560, L147 



\bibitem[Kubota \& Makishima(2004)]{Kubota_2004} Kubota, A., \& Makishima, K.\ 2004, \apj, 601, 428 


\bibitem[{{Lamb} \& {Sanford}(1979)}]{COMPLS}
{Lamb}, P. \& {Sanford}, P.~W. 1979, \mnras, 188, 555



\bibitem[{{Li} {et~al.}(2005){Li}, {Zimmerman}, {Narayan}, \&
  {McClintock}}]{KERRBB}
{Li}, L.-X., {Zimmerman}, E.~R., {Narayan}, R., \& {McClintock}, J.~E. 2005,
  \apjs, 157, 335

\bibitem[{{Liu} {et~al.}(2008){Liu}, {McClintock}, {Narayan}, {Davis}, \&
  {Orosz}}]{spin_m33}
{Liu}, J., {McClintock}, J.~E., {Narayan}, R., {Davis}, S.~W., \& {Orosz},
  J.~A. 2008, \apjl, 679, L37


\bibitem[{{McClintock} \& {Remillard}(2006)}]{MR06} {McClintock},
J.~E., \& {Remillard}, R.~A. 2006, in Compact Stellar X-ray Sources,
ed.\ W.\ Lewin \& M.\ van der Klis (Cambridge: Cambridge Univ.\ Press),
157

\bibitem[{{McClintock} {et~al.}(2007{\natexlab{b}}){McClintock},
  {Remillard}, {Rupen}, {Torres}, {Steeghs}, {Levine}, \&
  {Orosz}}]{JEM_H1743_2007} {McClintock}, J.~E., {Remillard}, R.~A.,
  {Rupen}, M.~P., {Torres}, M.~A.~P., {Steeghs}, D., {Levine}, A.~M., \&
  {Orosz}, J.~A. 2007{\natexlab{b}}, arXiv:0705.1034v1 [astro-ph]

\bibitem[{{McClintock} {et~al.}(2006){McClintock}, {Shafee}, {Narayan},
  {Remillard}, {Davis}, \& {Li}}]{spin_1915} {McClintock}, J.~E.,
  {Shafee}, R., {Narayan}, R., {Remillard}, R.~A., {Davis}, S.~W., \&
  {Li}, L.-X. 2006, \apj, 652, 518



\bibitem[{{Miller} {et~al.}(2008){Miller}, {Reynolds}, {Fabian}, {Cackett},
  {Miniutti}, {Raymond}, {Steeghs}, {Reis}, \& {Homan}}]{Miller_GX339}
{Miller}, J.~M., {Reynolds}, C.~S., {Fabian}, A.~C., {Cackett}, E.~M.,
  {Miniutti}, G., {Raymond}, J., {Steeghs}, D., {Reis}, R., \& {Homan}, J.
  2008, \apjl, 679, L113

\bibitem[{{Mitsuda} {et~al.}(1984){Mitsuda}, {Inoue}, {Koyama},
  {Makishima}, {Matsuoka}, {Ogawara}, {Suzuki}, {Tanaka}, {Shibazaki},
  \& {Hirano}}]{DISKBB} {Mitsuda}, K., {Inoue}, H., {Koyama}, K.,
  {Makishima}, K., {Matsuoka}, M., {Ogawara}, Y., {Suzuki}, K.,
  {Tanaka}, Y., {Shibazaki}, N., \& {Hirano}, T.  1984, \pasj, 36, 741



\bibitem[{{Nishimura} {et~al.}(1986){Nishimura}, {Mitsuda}, \&
{Itoh}}]{COMPBB} {Nishimura}, J., {Mitsuda}, K., \& {Itoh}, M. 1986,
\pasj, 38, 819

\bibitem[{{Orosz}(2003)}]{Orosz_inventory}
{Orosz}, J.~A. 2003, in IAU Symposium, Vol. 212, A Massive Star Odyssey: From
  Main Sequence to Supernova, ed. K.~{van der Hucht}, A.~{Herrero}, \&
  C.~{Esteban} (San Francisco: ASP), 365

\bibitem[{{Page} {et~al.}(2003){Page}, {Soria}, {Wu}, {Mason}, {Cordova}, \&
  {Priedhorsky}}]{Page_2003}
{Page}, M.~J., {Soria}, R., {Wu}, K., {Mason}, K.~O., {Cordova}, F.~A., \&
  {Priedhorsky}, W.~C. 2003, \mnras, 345, 639

\bibitem[{{Poutanen} \& {Svensson}(1996)}]{COMPPS}
{Poutanen}, J. \& {Svensson}, R. 1996, \apj, 470, 249

\bibitem[{{Remillard} \& {McClintock}(2006)}]{RR_JEM_review_2006}
{Remillard}, R.~A. \& {McClintock}, J.~E. 2006, \araa, 44, 49

\bibitem[{{Remillard} {et~al.}(2006){Remillard}, {McClintock}, {Orosz}, \&
  {Levine}}]{RR_H1743_2006}
{Remillard}, R.~A., {McClintock}, J.~E., {Orosz}, J.~A., \& {Levine}, A.~M.
  2006, \apj, 637, 1002





\bibitem[{{Rybicki} \& {Lightman}(1979)}]{Rybicki_Lightman} {Rybicki},
G.~B. \& {Lightman}, A.~P. 1979, {Radiative processes in astrophysics}
(New York, Wiley-Interscience)

\bibitem[{{Shafee} {et~al.}(2006){Shafee}, {McClintock}, {Narayan}, {Davis},
  {Li}, \& {Remillard}}]{Shafee_spin}
{Shafee}, R., {McClintock}, J.~E., {Narayan}, R., {Davis}, S.~W., {Li}, L.-X.,
  \& {Remillard}, R.~A. 2006, \apjl, 636, L113


\bibitem[{Shapiro}, {Lightman} \& {Eardley}(1976)]{Shapiro_1976}
{Shapiro}, S. L., {Lightman}, A. P., \& {Eardley}, D. M. 1976, \apj, 204, 187


\bibitem[{{Shrader} \& {Titarchuk}(1998)}]{Shrader_1998}
{Shrader}, C. \& {Titarchuk}, L. 1998, \apjl, 499, L31

\bibitem[{{Shrader} \& {Titarchuk}(1999)}]{Shrader_1999}
{Shrader}, C.~R. \& {Titarchuk}, L. 1999, \apjl, 521, L121

\bibitem[{{Swank}(1999)}]{RXTE}
{Swank}, J.~H. 1999, Nuclear Physics B Proceedings Supplements, 69, 12

\bibitem[{{Sunyaev} \& {Titarchuk}(1980)}]{Sunyaev_1980}
{Sunyaev}, R.~A. \& {Titarchuk}, L.~G. 1980, A\&A, 86, 121

\bibitem[{{Tanaka} \& {Lewin}(1995)}]{Tanaka_1995}
{Tanaka}, Y. \& {Lewin}, W.~H.~G. 1995, in X-ray Binaries, ed.\ W.\
Lewin, J.\ van Pardijs, \& E. van den Heuvel (Cambridge: Cambridge
Univ.\ Press), 126

\bibitem[{{Titarchuk}(1994)}]{COMPTT}
{Titarchuk}, L. 1994, \apj, 434, 570

\bibitem[{{Titarchuk} \& {Lyubarskij}(1995)}]{Titarchuk_95}
{Titarchuk}, L. \& {Lyubarskij}, Y. 1995, \apj, 450, 876

\bibitem[{{Titarchuk} {et~al.}(1997){Titarchuk}, {Mastichiadis}, \&
  {Kylafis}}]{BMC}
{Titarchuk}, L., {Mastichiadis}, A., \& {Kylafis}, N.~D. 1997, \apj, 487, 834

\bibitem[{{White} \& {Holt}(1982)}]{White_1982}
{White}, N.~E. \& {Holt}, S.~S. 1982, \apj, 257, 318

\bibitem[{{White} {et~al.}(1995){White}, {Nagase}, \&
{Parmar}}]{White_1995} {White}, N.~E., {Nagase}, F., \& {Parmar},
A.~N. 1995, in X-ray Binaries, ed.\ W.\ Lewin, J.\ van Pardijs, \&
E. van den Heuvel (Cambridge: Cambridge Univ.\ Press), 1

\bibitem[{{Yao} {et~al.}(2005){Yao}, {Wang}, \& {Nan Zhang}}]{Yao_2005}
{Yao}, Y., {Wang}, Q.~D., \& {Nan Zhang}, S. 2005, \mnras, 362, 229

\bibitem[Zdziarski et al.(2002)]{Zdziarski_2002}
Zdziarski, A. A., Leighly, K. M., Matsuoka, M., Cappi, M., \&
Mihara, T. 2002, \apj, 573, 505

\bibitem[Zhang et al.(1997)]{Zhang_1997} Zhang, S.~N., Cui, W., 
\& Chen, W.\ 1997, \apjl, 482, L155 

\bibitem[{{{\.Z}ycki} {et~al.}(1999){{\.Z}ycki}, {Done}, \& {Smith}}]{THCOMP}
{{\.Z}ycki}, P.~T., {Done}, C., \& {Smith}, D.~A. 1999, \mnras, 309, 561




\end{thebibliography}


  \begin{deluxetable}{lccccccccccc} 
  \rotate 
  \tabletypesize{\scriptsize} 
  \tablecolumns{      14}
  \tablewidth{0pc}  
  \tablecaption{Results of Fitting a Simulated {\sc compTT} Spectrum} 
  \tablehead{\colhead{MODEL} & & \colhead{$\rchi$/$\nu$} & \colhead{$\nh$} & \colhead{$\Gamma$}& \colhead{$\fsc$} & \colhead{${\rm Norm(PL)}$\tablenotemark{a}} & \colhead{$kT_0$}& \colhead{Norm\tablenotemark{b}} & \colhead{$kT_e$}& \colhead{$\tau_c$}\\ 
   &  & &           \colhead{(10$^{22}{\rm cm}^{-2}$)}  &  &     &             
 & \colhead{(keV)} & & \colhead{(keV)}}
  
\startdata 
{\sc compTT}\tablenotemark{c} & & \nodata & 0.1 & \nodata & \nodata & \nodata & 1. & 0.001 & 40. & 2. \\
\tableline



{\sc simpl-1 $\otimes$ bb} & & 1.00/731 & $0.28 \pm 0.01$ & $1.41 \pm 0.02$ & $0.84 \pm 0.01$ & \nodata &  $1.142 \pm 0.015$ & $10.9 \pm 0.4$ & \nodata & \nodata \\ 
{\sc simpl-2 $\otimes$ bb} & & 1.06/731 & $0.31 \pm 0.01$ & $1.37 \pm 0.02$ & $0.87 \pm 0.01$ & \nodata &  $1.292 \pm 0.010$ & $7.8 \pm 0.3$ & \nodata & \nodata \\ 
{\sc compbb}               & & 1.05/731 & $0.31 \pm 0.01$ & \nodata & \nodata & \nodata                 &  $1.292 \pm 0.010$ & $19.7 \pm 0.7$ & $43.6 \pm 2.2$ & $2.21 \pm 0.03$ \\
{\sc bb+powerlaw}          & & 2.02/731 & $0.68 \pm 0.01$ & $1.00 \pm 0.01$ & \nodata & $(5.0 \pm 0.2) \times 10^{-3}$ & $1.700 \pm 0.008$ & $0.89 \pm 0.02$ & \nodata & \nodata \\
\enddata

\tablenotetext{a}{{\sc powerlaw} normalization given at 1 keV in ${\rm photons \; s^{-1}cm^{-2}\,\keV^{-1}}$.}
\tablenotetext{b}{{\sc bb} and {\sc compbb} normalization $= \left(\frac{R/{\rm km}}{D/10\; \kpc}\right)^2$ for a blackbody of radius $R$ at a distance $D$; {\sc compTT} normalization is undefined.}
\tablenotetext{c}{{\sc compTT} model set to disk geometry (geometry switch = 1).}

\label{tab:COMPTT}
\end{deluxetable}

  \begin{deluxetable}{cccccccccccccccccccccccc} 
  \rotate 
  \tabletypesize{\scriptsize} 
  \tablecolumns{      23}
  \tablewidth{0pc}  
  \tablecaption{Spectral Fit Results}
  \tablehead{& & & & &  \multicolumn{1}{c}{{\sc phabs}}  &  &  \multicolumn{3}{c}{{\sc simpl}} & &  \multicolumn{2}{c}{{\sc diskbb}}  &  &  \multicolumn{2}{c}{{\sc powerlaw}}  \\ 
    \cline{6-6} \cline{8-10} \cline{12-13} \cline{15-16}  \\
    \colhead{Source} & \colhead{Mission} & \colhead{MJD} & \colhead{$\chi ^2 _\nu $/$\nu$} & \colhead{$\frac{L_{\scriptsize{\rm disk}}}{L_{\scriptsize{\textrm{Edd}}}}$\tablenotemark{a}} & 
  \colhead{$\nh$}  &  &  \colhead{Ver.\tablenotemark{b}} & \colhead{$\Gamma$} & \colhead{$f_{\scriptsize{\rm SC}}$} & & \colhead{$kT_{*}$}  &  \colhead{Norm\tablenotemark{c}}  &  &  \colhead{$\Gamma$}  &
    \colhead{${\rm Norm(PL)}$\tablenotemark{d}} \\ 
   \colhead{State} &  \colhead{Detector} & & & &  \colhead{(10$^{22}{\cm}^{-2}$)}  &  &  & & & &  \colhead{(keV)} & &  &&   } 
  \startdata  \label{tab.properties} 

H1743 & {\it RXTE} & 52797.6 &  0.64/44 &   0.19  & $ 1.94 \pm 0.17 $ & &\nodata &  \nodata &  \nodata & & $ 1.189 \pm 0.011 $ & $  568 \pm 31  $ & & $ 2.64 \pm 0.02 $ & $ 9.92 \pm 0.54 $ \\
              SPL & PCA & &     0.64/44 &   0.26 & $ 1.16 \pm 0.17 $ && S1 & $ 2.65 \pm 0.02 $ & $ 0.170 \pm 0.004 $ & & $ 1.157 \pm 0.011 $ & $  875 \pm 49 $  & & \nodata & \nodata \\
                        &&&     0.64/44 &   0.27  & $ 1.19 \pm 0.17$ && S2 & $ 2.65 \pm 0.02 $ & $ 0.224 \pm 0.005 $ & & $ 1.162 \pm 0.011 $ & $  878 \pm 50 $  & & \nodata & \nodata \\

\tableline
&\\

H1743 & {\it RXTE} & 52811.5 &  0.67/44 &   0.22 &   $ 1.53 \pm 0.15 $ & &\nodata &  \nodata &  \nodata & & $ 1.106 \pm 0.006 $ & $  869 \pm 33  $ & & $ 1.98 \pm 0.03 $ & $ 0.52 \pm 0.05 $ \\
               TD & PCA  & &    0.67/44 &   0.23 &   $ 1.48 \pm 0.15 $ && S1 & $ 1.98 \pm 0.03 $ & $ 0.030 \pm 0.001 $ & & $ 1.104 \pm 0.006 $ & $  909 \pm 35 $  & & \nodata & \nodata \\
                         &&&    0.67/44 &   0.23 &   $ 1.48 \pm 0.15 $ && S2 & $ 1.98 \pm 0.03 $ & $ 0.037 \pm 0.001 $ & & $ 1.105 \pm 0.006 $ & $  910 \pm 35 $  & & \nodata & \nodata \\

&\\
\tableline
&\\

LMC X--3 & {\it BeppoSAX}\tablenotemark{e} & 50415.5 &      1.05/729 &   0.58 &   $ 0.073 \pm 0.008 $ & &\nodata &  \nodata &  \nodata & & $ 1.279 \pm 0.011 $ & $ 24.5   \pm 0.8  $ & & $ 2.19 \pm 0.11 $ & $ 0.055 \pm 0.010$ \\
TD  & LECS,MECS, &   &     1.08/729 &   0.60 &  $ 0.044 \pm 0.003 $ && S1 & $ 2.41 \pm 0.45 $ & $ 0.062 \pm 0.021 $ & & $ 1.238 \pm 0.013 $ & $ 30.4 \pm 1.2 $  & & \nodata & \nodata \\
              & PDS &&     1.08/729 &   0.59 &  $ 0.044 \pm 0.003 $ && S2 & $ 2.46 \pm 0.48 $ & $ 0.085 \pm 0.033 $ & & $ 1.239 \pm 0.012 $ & $ 30.3 \pm 1.1 $  & & \nodata & \nodata \\

\enddata

\tablenotetext{a}{Bolometric ($0.1 - 20\;\keV$) luminosity of the disk component in Eddington units.  For H1743, we adopt nominal values: $M=10\;\msun$, $D=7.5$~kpc, and $i=70^\circ$.  The fiducial values used for LMC X--3 are $M=7.5\;\msun$ and $i=67^\circ$ \citep{Cowley_1983,Orosz_inventory}. For fits using {\sc simpl}, this quantity describes the seed spectral luminosity.}
\tablenotetext{b}{Version of {\sc simpl} being used, i.e., S1 for {\sc simpl-1} and S2 for {\sc simpl-2}.}
\tablenotetext{c}{For an accretion disk inclined by $i$ to the line of sight, with inner radius ${ R_{\rm in}}$ at distance ${D}$, ${\rm Norm} = \left(\frac{{ R_{\rm in}/\km}}{{ D/10\; \kpc}}\right) ^2\;\cos i $.}
\tablenotetext{d}{{\sc powerlaw} normalization given at 1 keV in ${\rm photons \; s^{-1}cm^{-2}\,keV^{-1}}$.}
\tablenotetext{e}{The cross-normalizations for $C_{\rm LM} \equiv {\rm LECS/MECS}$ and $C_{\rm PM} \equiv {\rm PDS/MECS}$ are fitted from $0.7-1$ and $0.77-0.93$ respectively. $C_{\rm LM}=0.802 \pm 0.283, 0.814 \pm 0.008, 0.813 \pm 0.008$ for the fits with {\sc powerlaw}, {\sc simpl-1}, and {\sc simpl-2}.  $C_{\rm PM}$ is pegged at 0.93 for the same fits.}

\tablecomments{All errors are presumed Gaussian and quoted at 1$\sigma$.}

\label{tab:H1743}
\end{deluxetable}

\clearpage

\begin{figure}
\plotone{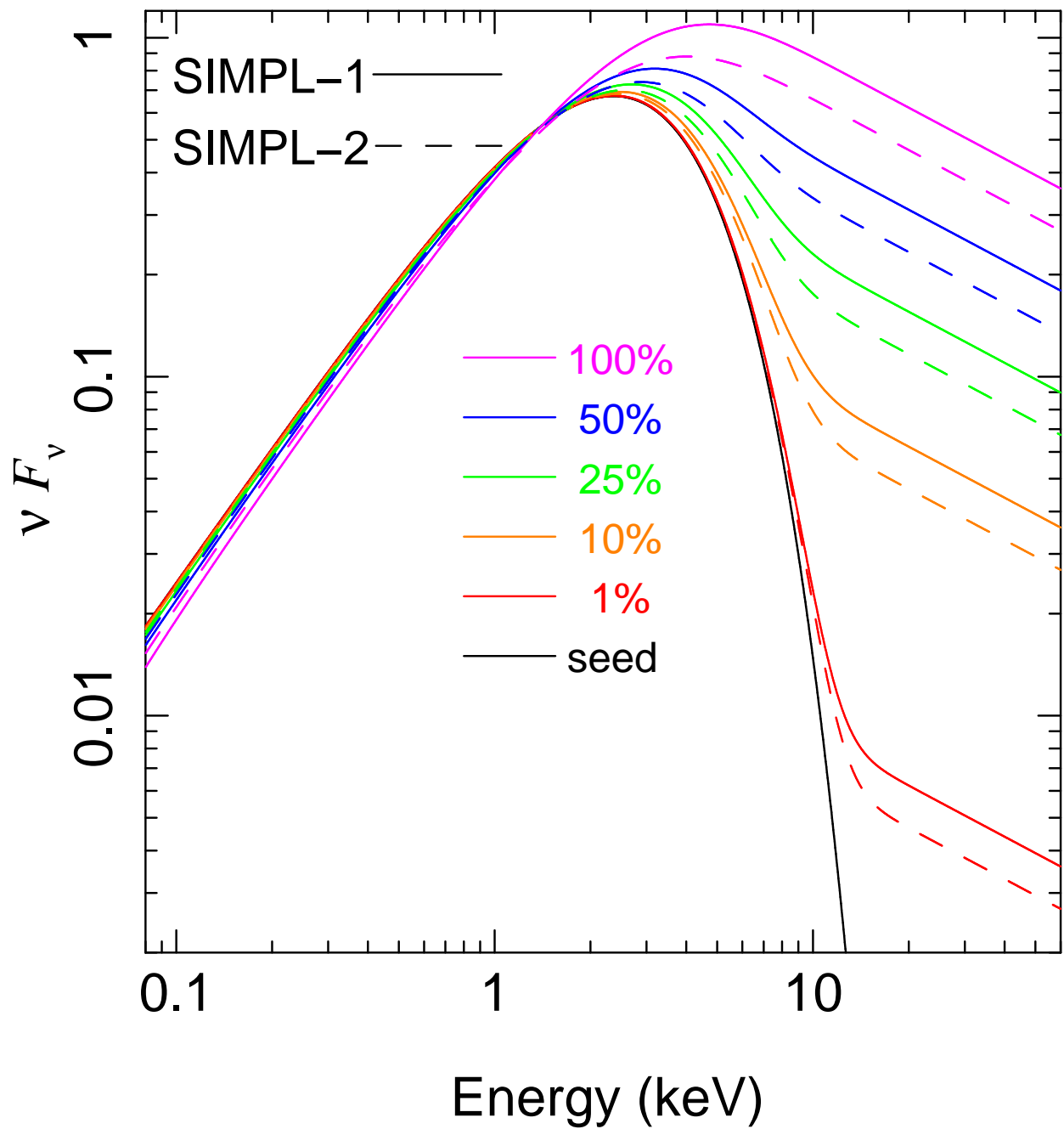}
\caption{Spectral energy density vs.\ photon energy for a sample
spectrum calculated with {\sc simpl-{\small 1}} (solid lines) and {\sc
simpl-{\small 2}} (dashed lines).  The models conserve photons and
Comptonize a seed spectrum, which in the case shown is {\sc diskbb}
with $kT_*=1 \; \keV$ (black line).  Ascending colored lines show
increasing levels of scattering, from $\fsc = 1-100$\%.}
\end{figure}

\begin{figure}
\plotone{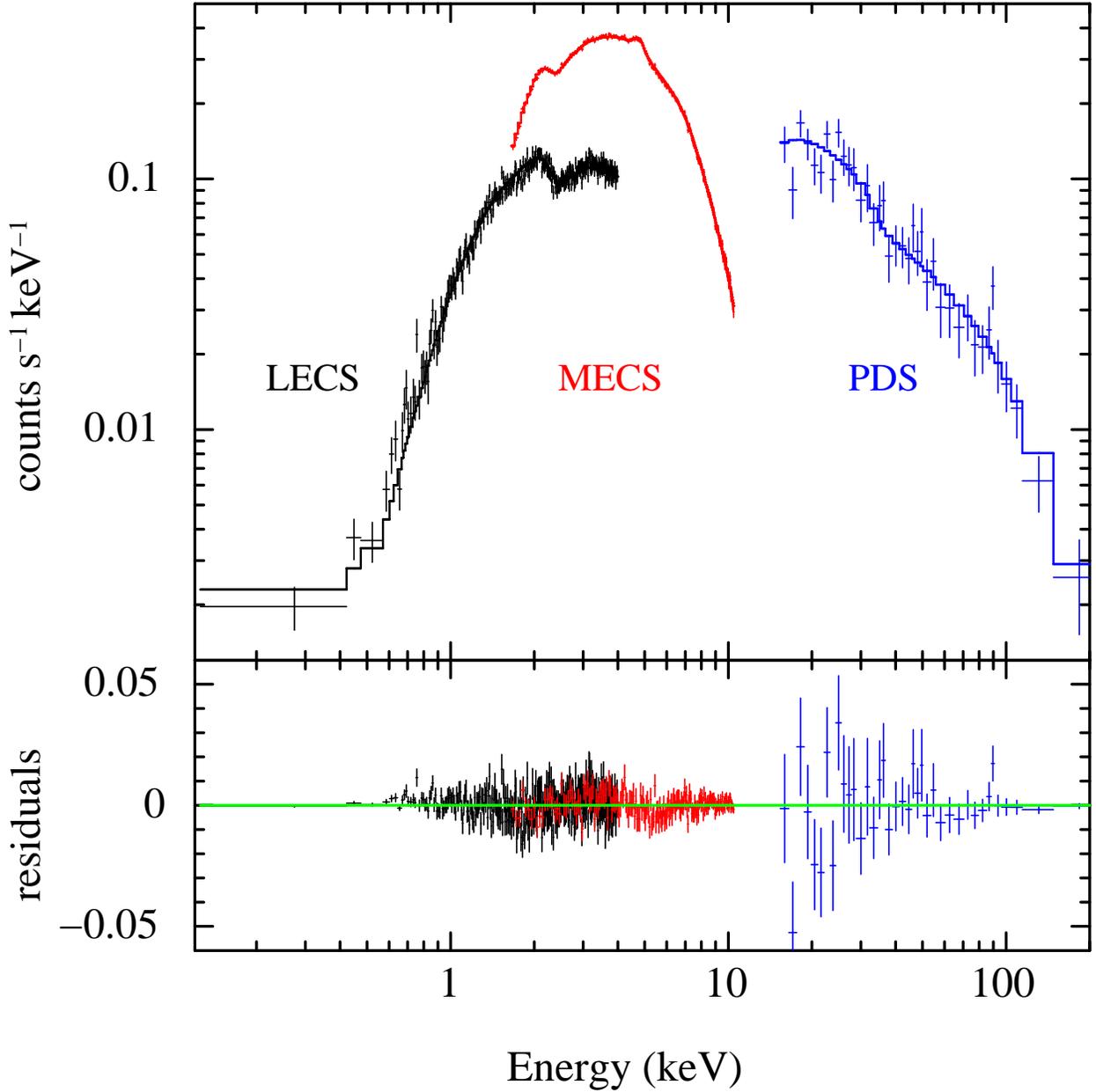} 
\caption{The data correspond to a simulated {\it BeppoSAX} observation
with a total of 2.1$\times 10^6$ counts; the spectrum was generated
using {\sc compTT}.  The histogram shows the fit achieved using {\sc
simpl-{\small 1}}.  This fit is performed over the recommended energy
ranges of the narrow-field instruments (NFI), as given by the Cookbook
for {\it BeppoSAX} NFI Spectral Analysis, yielding $\chi^2_\nu=1.00$.
For details, see Table~1.  This example demonstrates the ability of
{\sc simpl} to match a representative spectrum generated by a physical
model of Comptonization.}
\end{figure}

\begin{figure}
\plotone{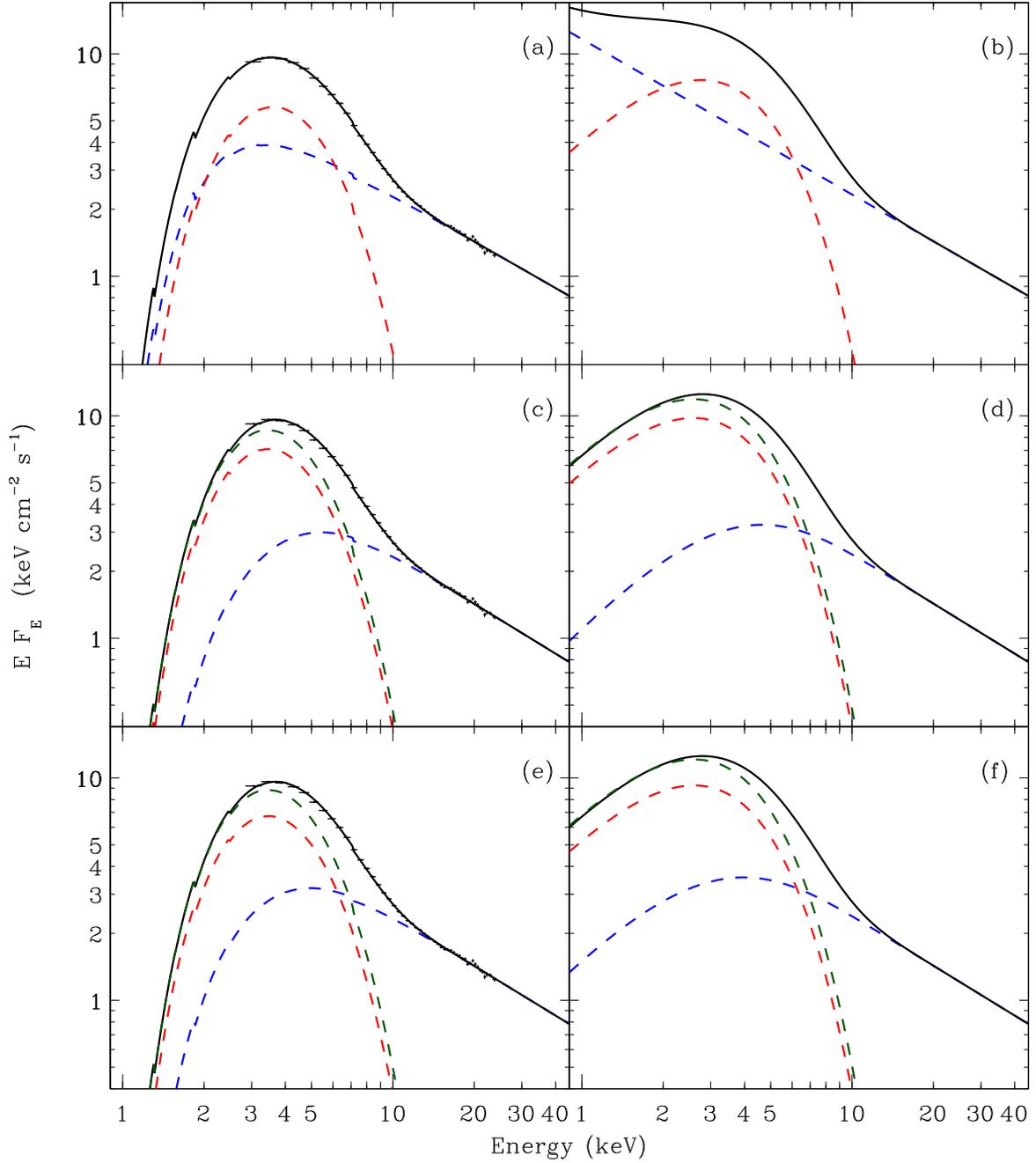}
\caption{{\it left:} Unfolded spectral fits to an {\it RXTE}
observation of H1743 in the SPL state and {\it right:} the corresponding
unabsorbed models.  Data are fitted using {\it (a,b):} {\sc
phabs$\times$(diskbb+powerlaw)}, {\it (c,d):} {\sc
phabs$\times$(simpl-{\small 1}$\otimes$diskbb)}, {\it (e,f):} {\sc
phabs$\times$(simpl-{\small 2}$\otimes$diskbb)}.  The composite model is
represented by a solid black line and the emergent disk and Compton
components are shown as red and blue dashed lines respectively.  The
seed spectrum for {\sc simpl} is shown (dashed) in green.  Contrasting
behaviors between {\sc simpl} and {\sc powerlaw} are most clearly
revealed in the unabsorbed models at low energies.  Spectral parameters
are given in Table~2.}
\end{figure}

\begin{figure}
\plotone{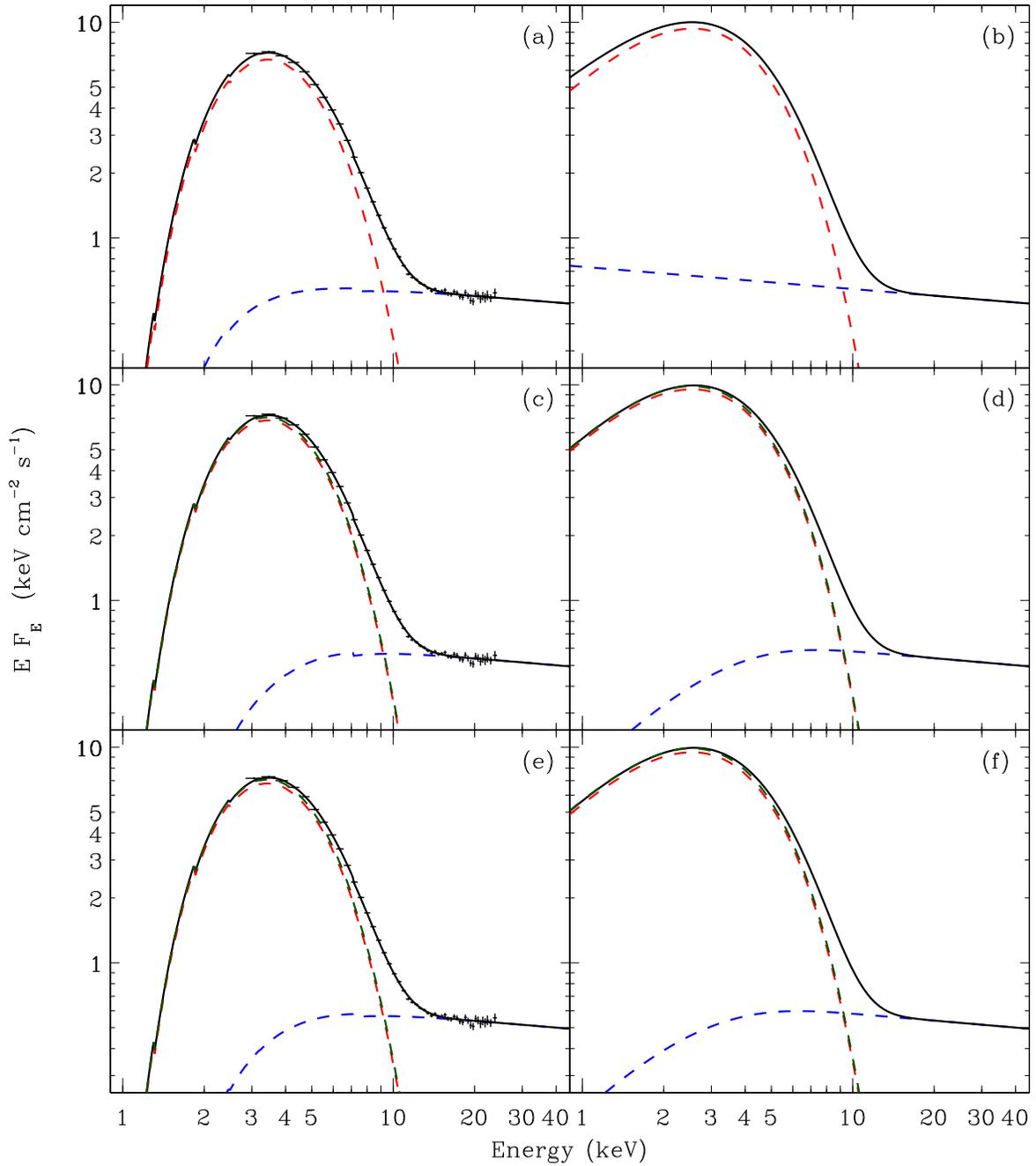}
\caption{Same as Figure~3 except that the results shown here are for an
{\it RXTE} observation of H1743 in the TD state.  The systematic
differences between the {\sc simpl} and {\sc powerlaw} fits are greatly
reduced compared to the differences shown for the SPL example in
Figure~3.}
\end{figure}

\begin{figure}
\plotone{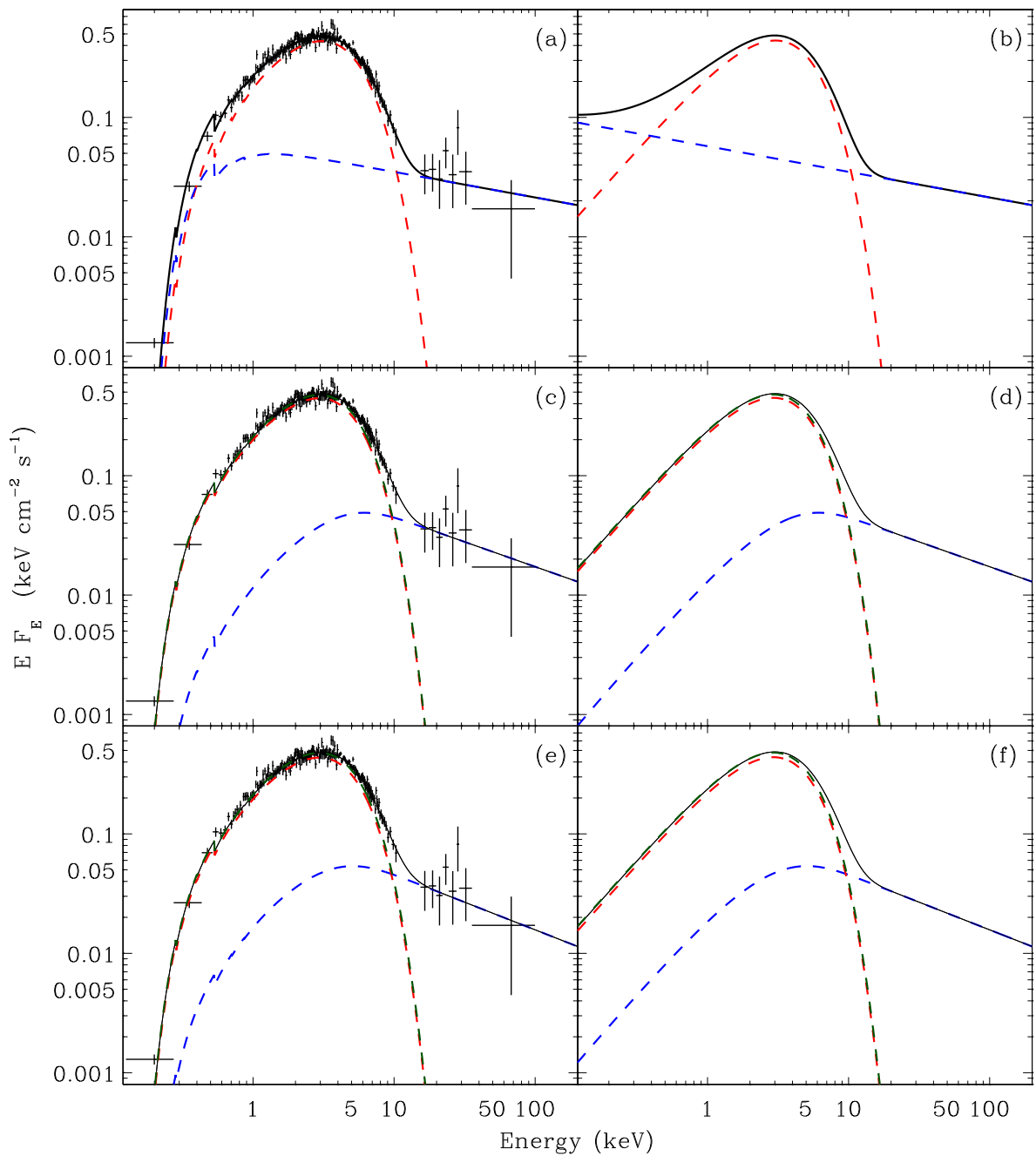}
\caption{Same as Figures 3 and 4 for a TD {\it BeppoSAX} spectrum of LMC
X--3.  The data have been rebinned for plotting purposes only and both
LECS and PDS counts have been rescaled by the fitted normalizations
given in Table~2.  At low energies (below $\sim$0.5 keV),
the unabsorbed model is strongly compromised for fits with {\sc
powerlaw}}.
\end{figure}

\end{document}